\documentclass[rmp,aps,onecolumn,nofootinbib]{revtex4}

\usepackage{amsmath}
\usepackage{tabularx}
\usepackage{palatino} 
\usepackage[]{graphicx}
\usepackage[]{setspace} 

\usepackage{epsfig}
\usepackage{epstopdf}

\begin{document}
\begin{spacing}{1.5}

\title{A Light Discussion and Derivation of Entropy}
\date{\today, version 1.01}
\author{Jonathon Shlens} 
\email{jonathon.shlens@gmail.com}
\affiliation{
Google Research \\
Mountain View, CA 94043}
\begin{abstract}
The expression for entropy sometimes appears mysterious -- as it often is asserted without justification. This short manuscript contains a discussion of the underlying assumptions behind entropy as well as simple derivation of this ubiquitous quantity.
\end{abstract}
\maketitle

The uncertainty in a set of discrete outcomes is the entropy.  In some text books an explanation for this assertion is often another assertion: the entropy is the average minimum number of yes-no questions necessary to identify an item randomly drawn from a known, discrete probability distribution. It would be preferable to avoid these assertions and search for the heart of the matter - \emph{where does entropy arise from?} This manuscript addresses this question by deriving from three simple postulates an expression for entropy.
\addtolength{\parskip}{\baselineskip}  

To gain some intuition for these postulates, we discuss the quintessential thought experiment: the uncertainty of rolling a die. How much uncertainty exists in the role of a die?  It is not hard to think of some simple intuitions which influence the level of uncertainty.

{\bf Postulate \#1}. \;\; A larger number of potential outcomes have larger uncertainty.

The more number of sides on a die, the harder it is to predict a role and hence the greater the uncertainty. Or conversely, there exists no uncertainty in rolling a single-sided die (a marble?). More precisely, this postulate requires that uncertainty grows {\it monotonically} with the number of potential outcomes.

{\bf Postulate \#2}.\;\; The relative likelihood of each outcome determines the uncertainty.

For example, a die which roles a $\mathtt{6}$ a majority of the time, contains less uncertainty than a standard, unbiased die. The second postulate goes a long way because we can express the uncertainty $H$ as a function of the probability distribution $p = \{p_1, p_2, \ldots, p_A \}$ dictating the frequency of all $A$ outcomes or, in short-hand, $H[p]$. Thus, by the first postulate $\frac{dH}{dA} > 0$ since the uncertainty grows monotonically as the number of outcomes increases. Strictly speaking, in order for the derivative to be positive, the derivative must exist in the first place, thus we additionally assume that $H$ is a {\it continuous} function.

{\bf Postulate \#3}\;\; The {\it weighted} uncertainty of independent events must sum.

This final postulate was a stroke of genius recognized by Claude Shannon~\cite{Shannon1949}.\footnote{Ironically, this idea is so central that it is sometimes overlooked~\cite{DeWeese1999}.} In the case of rolling two dice, this means that the total uncertainty of rolling two independent dice must equal the sum of the uncertainties for each die alone.  In other words, if the uncertainty of each die is $H_1$ and $H_2$ respectively, then the total uncertainty of rolling both die simultaneously must be $H_1 + H_2$.

\begin{figure}
\centerline{\psfig{file=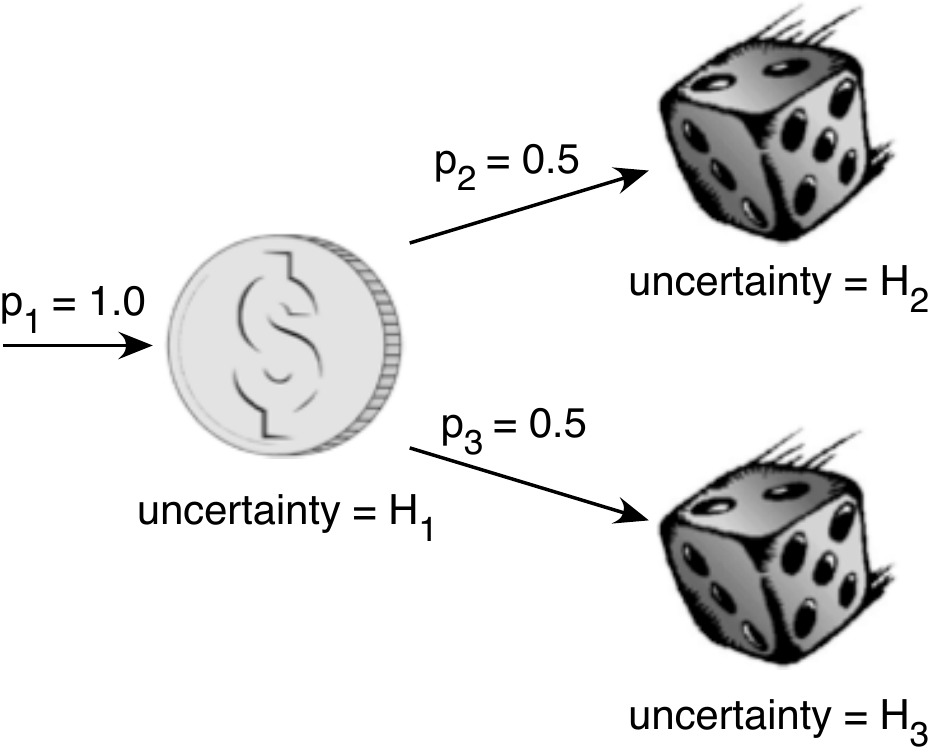,width=0.45\textwidth}}
\caption{A simple example of the composition rule. Imagine we have a 2-sided coin and two 6-sided dice, where the first role of the coin determines which die we will roll. The total uncertainty of this operation is the sum of the uncertainties of each object weighted by the probability of the action. In a single operation we always flip the coin so $p_1 = 1$, but the probability of rolling each die is determined by the coin, $p_2 = 0.5$ and $p_3 = 0.5$, respectively. Thus, the total uncertainty of the operation is $\sum_{i}{p_i H_i}$.}
\label{fig:tree-postulate}
\end{figure}

{\it Weighted} refers to the fact that the uncertainties should be weighted by the probability of occurence. In the two die example both die are rolled but what if there exist a probability of the role itself? This notion is sometimes referred to as the {\it composition rule} and is best understood by examining Figure~\ref{fig:tree-postulate}.

Shannon proved that these simple postulates lead to a unique mathematical expression for uncertainty (see Appendix~\ref{sec:entropy-derivation}).  For the probability distribution $p$ the only function that matches these intuitions is
\begin{equation}
H[p] \equiv -\sum_{i=1}^{A}{p_i \log_2 p_i}.
\label{eq:entropy-definition}
\end{equation}
$H$ is termed the {\it entropy} of the distribution and is the same quantity observed in physics and chemistry (with different units)~\cite{Jaynes1957a, Jaynes1957b, Brillouin2004}. Note that $x \log x \equiv 0$ because we attribute zero uncertainty to impossible outcomes. $H$ measures our definition of uncertainty as specified by the three previous statements and is often viewed as a measure of variability or concentration in a probability distribution. The appendix contains a simple derivation of Equation~\ref{eq:entropy-definition} following solely from the three postulates.
\addtolength{\parskip}{-\baselineskip}  

\appendix
\section{Derivation}
\label{sec:entropy-derivation}

We derive the entropy, Equation~\ref{eq:entropy-definition}, following the original derivation of Shannon~\cite{Shannon1949} solely using the three postulates in the previous description (see also~\textcite{Carter2000}). The strategy for deriving the entropy consists of two parts: (1) the specific case of a uniform  distribution, (2) the general case of a non-uniform distribution.
 \addtolength{\parskip}{\baselineskip}  
  
We begin with the {\it composition rule}. Consider two separate, independent, uniform probability distributions with $x$ and $y$ elements respectively. The composition law requires that
$$H(x) + H(y) = H(xy).$$
where $H(x)$ refers to the entropy of a uniform distribution with $x$ outcomes. Intuitively, this is equivalent to saying that the uncertainty of simultaneously rolling a $x$-sided and a $y$-sided die is equal to the sum of the uncertainties for each die alone. In the single die case there exist $x$ and $y$ equally probable outcomes respectively, and in the simultaneous case, there exist $xy$ equally probable outcomes. To derive the uniform form for $H$ we take the derivative with respect to each variable:
\begin{eqnarray*}
\frac{dH(x)}{dx} & = & y \; \frac{dH(xy)}{dx} \\
\frac{dH(y)}{dy} & = & x \; \frac{dH(xy)}{dy}
\end{eqnarray*}
The variable names $x$ and $y$ are arbitrary thus $\frac{dH(xy)}{dx} = \frac{dH(xy)}{dy}$. Substituting one equation into another and a little algebra yields
$$x \frac{dH(x)}{dx} = y \frac{dH(y)}{dy}.$$
Each side of this equation is solely a function of an arbitrary choice of variables, thus equality can only hold for all $x$ and $y$ if and only if both sides equal a constant,
$$x \frac{dH(x)}{dx} = k$$
where $k$ is some unknown constant. Solving the above equation for $\frac{dH(x)}{dx}$ and integrating over $x$ yields $H(x) = k \log x + c$ where $c$ is another constant. We set $c=0$ because there is zero uncertainty when only a single outcome is possible. The first postulate requires that $\frac{dH(x)}{dx} > 0$ implying that $k>0$. The selection of the base of the logarithm can absorb the choice of the coefficient $k$ in front. We select base 2 logarithms to provide the familiar units of {\it bits}, resulting in the final form of the equation,
\begin{equation}
H(x) = \log x ,
\label{eqn:entropy-uniform}
\end{equation}
and completing the first section of the derivation.

The non-uniform case extends from the uniform case by assuming the probability of each outcome $p_i$ can be expressed as $p_i \equiv \frac{n_i}{N}$ where $n_i$ and $N = \sum_{i}{n_i}$ are integers. For example, if we had $N=10$ fruits but only $30\% $ are oranges, then $p_{orange} = 0.3$ and $n_{orange} = 3$. Thus, we assume that each probablity can be expressed as a fraction with an integer numerator and denominator. The uncertainty of the complete set of outcomes is $\log_2 N$ by Equation~\ref{eqn:entropy-uniform}. The composition rule requires that $\log_2 N$ is equal to the sum of:
\begin{enumerate}
\item the uncertainty of an item drawn from $p$ (e.g. any orange out of the fruit).
\item the {\it weighted} uncertainty of selecting an item uniformly from $n_i$ items (e.g. 1 orange out of all oranges)
\end{enumerate}
The first quantity is $H[p]$ and it is the quantity we wish to derive. The second quantity is
$$\sum_{i=1}^{A}{p_{i}H(n_i)} = \sum_{i=1}^{A}{p_{i} (\log_2 n_i}),$$
where $p_i$ weights the uncertainty associated with each outcome.  Putting this
altogether we get
\[\log N = H[p] + \sum_{i=1}^{A}{p_i (\log_2 n_i})\]
and with a little algebra,
\begin{eqnarray*}
H[p] & = & \log_2 N - \sum_{i=1}^{A}{p_i \log_2 n_i} \\
     & = & - \sum_{i=1}^{A}{p_i \log_2 \frac{n_i}{N}}
\end{eqnarray*}
Recognizing the definition of $p_i$ within the logarithm, we recover the definition of entropy (Equation~\ref{eq:entropy-definition}). Of course this does not hold if $p_i$ are irrational but nonetheless can be approximated to arbitrary accuracy for large enough $N$. Combined with the continuity assumption on $H$, this expression must likewise hold for irrational $p_i$.
\addtolength{\parskip}{-\baselineskip}  

\bibliographystyle{natbib}
\bibliography{entropy}

\end{spacing}
\end{document}